\documentclass[prd,superscriptaddress,preprintnumbers,amsmath,amssymb,nofootinbib,tightenlines,11pt]{revtex4}

\usepackage{amsmath}
\usepackage{amsfonts}
\usepackage{amssymb}
\usepackage{epsfig}
\usepackage{graphicx}
\usepackage{color}
\usepackage{slashed}
\usepackage{epstopdf}
\usepackage[utf8]{inputenc}

\usepackage{hyperref}

\newcommand{\be}{\begin{equation}}
\newcommand{\ee}{\end{equation}}
\newcommand{\bea}{\begin{eqnarray}}
\newcommand{\eea}{\end{eqnarray}}

\begin{document}

\preprint{}

\title {Towards coarse graining of discrete Lorentzian quantum gravity
}

\author{Astrid Eichhorn}
\email[]{a.eichhorn@thphys.uni-heidelberg.de} 
\affiliation{Institute for Theoretical Physics, University of Heidelberg, Philosophenweg 16, 69120 Heidelberg, Germany}

\begin{abstract}
A first step towards implementing a notion of coarse graining in an intrinsically Lorentzian, discrete quantum- gravity approach, namely causal set quantum gravity is taken. It makes use of an abstract notion of scale, based on counting  the number of discrete elements. To that end, the space of actions for causal set quantum gravity is written in a matrix-model-like language, and a flow equation for the effective action of the model is derived from the path integral. 
\end{abstract}


\maketitle

\section{Introduction}

A first step to develop coarse- graining tools in a discrete, Lorentzian approach to quantum gravity, namely causal sets \cite{Bombelli:1987aa,Sorkin:2003bx,Dowker:2005tz,Henson:2010aq,Surya:2011yh,Dowker:aza}, is taken. The underlying motivation is threefold:

First, bridging the gap between microscopic and macroscopic scales, and testing whether a smooth spacetime emerges from discrete fundamental building blocks, requires tools that allow to evaluate the path-integral over discrete configurations. In this note, a flow equation is set up, based on the idea to perform the path integral in steps to derive the effective action of the model.  In contrast to continuum Quantum Field Theories, where a stepwise evaluation of the path-integral is typically based on momentum shells, i.e., a local coarse graining, a discrete notion of scale, based on counting elements, is available in discrete settings \cite{Brezin:1992yc,Eichhorn:2013isa}, such as, e.g., causal sets and is made use of here.

A second motivation is the question of predictivity of quantum- gravity models featuring a fundamental discreteness scale that serves as a physical ultraviolet cutoff. Thinking about coarse graining in these models shines a spotlight on an open question:
as coarse-graining generically washes out microscopic details, one expects that a possibly infinite set of microscopic parameters can be introduced that  do not leave detectable imprints at large scales. Their presence could lead to a breakdown of predictivity at small scales. This is similar to the breakdown of predictivity  in an effective field theory  at the ``scale of new physics": there, infinitely many interactions with corresponding couplings are no longer suppressed and become important. In contrast, the absence of a fundamental physical cutoff, i.e., the requirement of a well-defined continuum limit, generically selects a finite dimensional hypersurface in the space of possible models that relates all but finitely many of the independent couplings  to each other, resulting in a finite number of remaining free parameters. A corresponding restriction to a predictive subspace of the infinite dimensional space of actions appears to be absent in the presence of a finite physical ultraviolet (UV) cutoff, possibly resulting in a model that is not predictive near the discreteness scale.
As the intuition behind this argument is based on local quantum field theories, but causal sets are nonlocal, it is crucial to formulate suitable coarse-graining tools that allow to explore the space of actions in this setting and confirm or refute the validity of the above argument.

A third motivation to set up coarse graining tools for discrete Lorentzian quantum gravity comes from a slightly unorthodox view on causal sets, based on the asymptotic-safety program  \cite{Weinberg:1980gg,Reuter:1996cp, ASgrav,Dona:2013qba}. Asymptotic safety is the existence of an interacting, ultraviolet fixed point of the Renormalization Group (RG) flow in a continuum quantum field theory of the metric.
Rephrased in a language adapted to a discrete setting, the interacting fixed point implies that a well-defined continuum limit exists, at which the discreteness scale can be taken to zero and microscopic details of the model do not play a role due to universality arising in the scaling regime underlying the fixed point. In the standard approach to search for asymptotic safety, one employs continuum RG tools that are adapted to a Euclidean setting. While it is possible to derive approximations  to the full RG flow   that admit a Wick-rotation to the Lorentzian setting \cite{Manrique:2011jc,Rechenberger:2012dt,Biemans:2017zca}, studies of the fully Lorentzian path integral for quantum gravity do not (yet) exist. On the other hand, the potential impact of Lorentzian signature is highlighted by the differences in the phase diagrams of Euclidean and Dynamical Triangulations \cite{Ambjorn:1998xu,Ambjorn:2011cg,Ambjorn:2012ij,Laiho:2016nlp}. Thus it is desirable to approach the search for a fixed point from a new angle which is inherently Lorentzian. The usual interpretation of the discreteness scale in causal sets is as a physical, fundamental scale, but nothing in principle prohibits taking that scale to zero and exploring whether a well-defined continuum limit exists. This might require an additional restriction on the path-integral measure to discard non-manifoldlike configurations, if no dynamics can be found that suppresses these sufficiently, and could thus highlight which type of quantum fluctuations of gravity (e.g., including fluctuations in topology, dimensionality etc.) are compatible with asymptotic safety.

\section{Causal set quantum gravity}\label{sec:CSQG}

Causal set quantum gravity \cite{Bombelli:1987aa}, see \cite{Sorkin:2003bx,Dowker:2005tz,Henson:2010aq,Surya:2011yh,Dowker:aza} for reviews, is a Lorentzian model of quantum spacetime. It unifies the emphasis on causal structure with hints for physical discreteness in quantum gravity, such as, e.g., from black-hole entropy, and imposes the existence of a fundamental minimum scale -- understood as a minimum spacetime volume -- at the kinematical level \footnote{Note that a physical minimum scale need not necessarily be imposed at the kinematical level but can also emerge dynamically. For instance, asymptotically safe quantum gravity is a continuum quantum field theory for the metric, but nevertheless features a form of minimum length scale \cite{Lauscher:2005qz,Percacci:2010af}: Beyond the transition scale to the fixed point regime, physics becomes scale invariant and all dimensionful quantities scale according to their canonical dimensionality, i.e., the transition scale acts as a dynamically emergent minimum scale.}. The approach is based on the path-integral framework, where the configurations to be summed over are all possible causal sets. A causal set $\mathcal{C}$ is a partially ordered set, where the elements are to be interpreted as discrete ``atoms of spacetime" and the order relation $\prec$ is a causal order, i.e., for two elements $x,y$, $x\prec y$ is a causal relation between the two spacetime points, such that $x$ is in the causal past of $y$. Two elements  share a link if they are causally related and there is no other element between them, i.e., a link, denoted by $\prec \ast$, is an irreducible relation.
In order for this interpretation to work such that a physical spacetime can emerge from the discrete setting, $\prec$ should be acyclic, i.e., if $x \prec y$ and $y \prec x$, then $x=y$ and transitive, i.e., if $x \prec y$ and $y \prec z$ then $x \prec z$. Discreteness is achieved by demanding local finiteness, i.e., $\forall\,  x, y\, \in \mathcal{C}, {\rm card}:\{z|\, x \prec z \prec y \}< \infty$. A spacetime equipped with a metric might emerge from this setting, as underscored by a theorem by D.~Malament, \cite{Malament}, stating that if there is a bijective map between two spacetimes that are past and future distinguishing and their causal structure is preserved by the map, then that map is a conformal isomorphism. In other words, causal relations determine the metric up to a conformal factor. The conformal factor encodes the volume of spacetime, and in the discrete case can be recovered by \emph{counting}: Associating a fundamental volume (e.g., a Planck volume if the discreteness scale is the Planck scale) to each element of a causal set, the total volume is encoded in the number of elements. This underlies the causal set slogan, coined by R.~Sorkin,  that `order+ number = geometry' \cite{RSorkin1, RSorkin2}. \\
The set of all causal sets contains manifoldlike ones, i.e., causal sets that are approximated by a continuum spacetime, and non-manifoldlike ones. Those dominate the set of all causal sets entropically, i.e., picking a causal set at random is almost certain to yield a non-manifoldlike one, at least beyond a causal set size $N \approx 40$ \cite{Kleitman,Henson:2015fha}. The dynamics must counteract the entropic dominance of non-manifoldlike causal sets, such that a spacetime equipped with a metric can emerge from the path integral. Based on progress on the construction of a discrete d'Alembertian for scalar fields on a causal set \cite{Sorkin:2007qi,Johnston:2009fr,Dowker:2013vba,Aslanbeigi:2014zva}, the Benincasa-Dowker action \cite{Benincasa:2010ac,Benincasa:2010as}, corresponding to a discrete version of the Einstein action, is a proposal for a dynamics. Based thereupon, Monte Carlo simulations of the path sum for (restricted classes of) causal sets are possible \cite{Surya:2011du,Glaser:2017sbe}. In fact, it has recently been shown that this action leads to destructive interference of the non-manifoldlike class of two-level orders, a subleading class compared to the most dominant non-manifoldlike orders, but entropically favored over the manifoldlike causets \cite{Loomis:2017jhn}.

\section{Theory space for causal sets}\label{sec:theoryspace}
The Einstein-Hilbert action, or an appropriate discretization, is often chosen as the microscopic dynamics in discrete quantum- gravity settings. However, an analogy highlights reasons suggesting an alternative choice: the microscopic dynamics of condensed-matter systems are distinct from the effective dynamics of emergent low-energy excitations -- e.g., for graphene, a honeycomb lattice of carbon atoms, the electrons are non-relativistic, while the emergent effective electronic excitations obey a Dirac-type dispersion relation, i.e., the emergent and the microscopic dynamics are different, and do not even share the same symmetry.
Similarly, in gravity it is the low-energy effective degrees of freedom that should obey Einstein-Hilbert dynamics. In contrast, the microscopic dynamics could, or potentially even has to, differ. A similar reasoning motivates the analysis of analogue gravity models.
 In causal sets, the dynamics of single causal set elements or configurations is expected to be distinct from the emergent low-energy dynamics of excitations of spacetime. Thus, a priori, any ansatz could be made for the microscopic dynamics, which reduces to the Einstein-Hilbert dynamics for the emergent low-energy degrees of freedom. 
The connection between microscopic and effective dynamics can be established by a coarse-graining procedure. 
Under a coarse-graining step, the dynamics of a system changes, unless a form of scale invariance is realized. The change can typically be encoded in a scale-dependence of the couplings parameterizing the strength of different interactions, as well as in the generation of new interactions\footnote{Note that herein the notion of scale can be very abstract and need not be associated to a length or momentum scale. For instance, in matrix and tensor models for quantum gravity, an appropriate notion of scale is given by the number of degrees of freedom \cite{Brezin:1992yc,Eichhorn:2013isa,Eichhorn:2014xaa,Eichhorn:2017xhy}: The macroscopic, coarse-grained description has fewer degrees of freedom than the microscopic one. This notion of scale is useful for the case of causal sets.}.
In most cases, the ``evolution" of a system under a coarse-graining procedure explores the full theory space of the system. Theory space is the space of all couplings compatible with the symmetries of the model, i.e., each point in it defines a possible dynamics consistent with all symmetries at hand. Typically, this space is infinite dimensional, and causal sets are no exception \footnote{Strictly speaking, for finite-size causal sets, theory space should be finite, but scales with the causal set size $N$, which should be very large in order for the observable Hubble volume to be described by an underlying causal set model with a discreteness scale set by the Planck scale.}. 

\subsection{Translating causal sets into matrix-model language}
The theory space for causal sets is determined by a discrete analogue of diffeomorphism symmetry, namely label-independence. Although the elements in any given causal set can be labelled, the causal set dynamics must not depend on these labels, see, e.g., \cite{Rideout:1999ub}. Accordingly, the theory space of causal sets can only contain couplings related to quantities that can be counted, such as, e.g., the number of elements, the number of links, the number of causal intervals of a certain size, or the number of antichains. 
The complete information on a causal set, from which all these quantities can be derived, is encoded in its link matrix \cite{Johnston:2010su}, which is defined as follows
\be
L_{ij}=  \begin{cases}
    1, & \text{if $i\prec\ast j$}.\\
    0, & \text{otherwise}.
  \end{cases} 
\ee
Therefore, one can represent the path integral over all causal sets with the action $S$ in the form
\be
\int \mathcal{D}\mathcal{C}\,e^{i\,S[\mathcal{C}]} \rightarrow \int \mathcal{D}L_{ij}\,e^{i\,S_L[L_{ij}]}\,.
\ee
Here, one might introduce an additional weight factor that avoids counting several link matrices corresponding to the same causal set and differing only by a relabelling, i.e., one might gauge fix the discrete diffeomorphisms.
The most general dynamics can only depend on quantities such as the total number of links and $m$-paths, as no other label invariant information can exist. Herein, an $m$-path is defined as a set of elements $i_0\prec\ast i_1...\prec\ast i_m$, i.e., the cardinality of an $m$-path is $m+1$.
The number of $m$-paths  between two elements is encoded in powers of the link matrix, as
\be
\left(L^m\right)_{ij} = {\rm \#}\left(\,m-\,{\rm paths\, between}\, i\, {\rm and}\, j\right). 
\ee
Thus, the total number of $m$ paths is given by $\sum_{i,j} \left(L^m\right)_{ij}$. 

The most general action that depends on all elements of the link matrix is
\be
S(L) = \sum_{i,j}f_{ij}(L_{ij}),
\ee
where the $f_{ij}$ can be any function, and a priori not need to be the same function for different $i,j$.
On the other hand, discrete diffeomorphisms must leave the action invariant. A discrete diffeomorphism that consists, e.g., in exchanging the first two labels,  acts by exchanging the 1st and 2nd row with each other followed by an exchange of the 1st and 2nd  column. This implies
\be
f_{i2}=f_{i1}, \quad f_{2i}=f_{1i}, \mbox{ for } i \neq 1,2, \mbox{ and } f_{11}=f_{22}, \, f_{12}=f_{21}.
\ee
Continuing in the same fashion, one derives  that $f_{ij}$ must be independent of $i,j$ for $i\neq j$. For the diagonal elements, one obtains $f_{ii}= f_{jj}$, $\forall i,j$. However, the diagonal entries for all link matrices are zero, accordingly the dependence on the diagonal element only adds a link-matrix-independent constant to the action, which can be dropped. Thus one concludes that the dependence of the action on all elements of the link matrix must be the same, just as one would expect for a setting which respects discrete diffeomorphism symmetry and where accordingly no entry of the link matrix can be distinguished. If one assumes that one can Taylor expand that function, this yields
\be
S[L]= \sum_J \hat{g}_{J} \sum_{i,j}\left(L_{ij}\right)^J + \sum_m g_m \sum_{ij}\left(L^m\right)_{ij}.
\ee
For the first sum over $J$, since the entries of $L$ are zeros and ones only, $\left(L_{ij}\right)^J=L_{ij}$. Accordingly that sum collapses, and under the assumption that $S$ can be expanded, 
\be
S[L] = \sum_{m=1}^{\infty} g_m \sum_{i,j}\left(L^m \right)_{ij}.
\ee
For causal sets of finite size $N$, only $g_m,\, m<N$ contribute, as $L^m=0$ for $m>N$, since there cannot be a chain longer than $N$ elements.

 Note that the assumption that the action can be Taylor expanded is a strong one. In fact, it appears that already the discrete analogue of the Einstein action does not respect this assumption. Thus, theory space also contains couplings in addition to the $g_m$, which arise from functions that do not admit a Taylor expansion. Whether this theory space admits the introduction of a basis needs to be studied. For the setup of simple coarse-graining procedures, it might be sufficient to restrict the theory space to the $g_m$.
 
As an example, the two-dimensional Benincasa-Dowker action, as given in \cite{Benincasa:2010as} can be translated into this notation. It is given by
\be
S_{\rm BD}^{(2)}[\mathcal{C}]= N- 2N_1 +4N_2-2N_3,
\ee
where $N$ is the number of elements in $\mathcal{C}$ and $N_i$ is the number of inclusive order intervals of cardinality $i+1$. For $i=1$, these are the number of links, and for $i=2$, the 2-paths. For $i=3$, the inclusive order intervals include 3-paths as well as causal diamonds, i.e., subsets, where a top and a bottom element are related via two distinct 2-chains, and the two "intermediate" elements are unrelated. The inclusive order interval between two elements is a diamond if and only if the number of 2-paths between them is exactly two, and thus one can obtain the number of causal diamonds $N_D$ by
\be
N_D = \sum_{ij} \delta_{L^2_{ij}, 2}.
\ee
Accordingly, it holds that
\be
S_{\rm BD}^{(2)}[\mathcal{C}] = N - \sum_{ij}\left(2 L_{ij}-4 L^2_{ij}+2\left( L^3_{ij}+ \delta_{L^2_{ij},2}\right)\right).
\ee
Similarly, the 4-dimensional Benincasa- Dowker action, given by $
S_{\rm BD}^{(4)}[\mathcal{C}]= N-N_1+9 N_2-16 N_3+8 N_4$, 
can be rewritten as
\be
S_{\rm BD}^{(4)}[\mathcal{C}]= N - \sum_{ij}\left( L_{ij} -9L^2_{ij} +16 \left(L^3_{ij} +\delta_{L^2_{ij},2}\right)-8 \left(L^3_{ij}+ \delta_{L^2_{ij},3}+\delta_{L^2_{ij},1}\delta_{L^3_{ij},1}+\delta_{L^3_{ij},2} \right) \right).
\ee

To dampen fluctuations, the action is usually altered by the introduction of a ``non-locality" scale $l> l_{\rm Pl}$, where $l_{\rm Pl}$ is the discreteness scale, and $\epsilon = l_{Pl}^2/l^2 \in (0,1)$ \cite{Aslanbeigi:2014zva}. In the action, one sums over more ``layers" of elements, weighted by a function of $\epsilon$. This can be expressed as a function of $n$-element intervals which can be rewritten in terms of the link matrix.
\subsection{Predictive dynamics in the discrete quantum-gravity setting}\label{sec:pred}
The infinite dimensionality of theory space raises the general question how models can be predictive, i.e., fully specified in terms of a finite number of free parameters. This  requires infinitely many relations between the different couplings.
In continuum Quantum Field Theories, the infinite-dimensional theory space is the space of all couplings compatible with the symmetries and with any positive integer power of derivatives. There are two settings which yield physical predictions that only depend on a finite number of free parameters:\\
 In effective field theories which only hold up to a ``scale of new physics" $\Lambda$. There, each of the infinitely many couplings is a free parameter that is determined by the underlying, more microscopic theory. Without knowing the microscopic theory, one cannot determine the values of those couplings. On the other hand, there is an ordering principle in theory space, namely the mass-dimensionality of couplings. Those couplings associated to higher-order field-monomials are of increasingly negative mass dimensionality. The central assumption of effective field theory is that the mass-scale of new physics, $\Lambda$, sets the values of couplings, and their dimensionless values are all of order one. Hence, only a finite number of couplings matters for the value of a low-energy observable when it is determined at finite precision. Thus, at low energies, only finitely many couplings matter, and all others are suppressed, at least in a theory space based on locality.\\
The second predictive setting is that of a fundamental quantum field theory where the continuum limit $\Lambda \rightarrow \infty$ can be taken in a controlled way. Usually, this is achieved by approaching a fixed point of the Renormalization Group - more exotic possibilities like limit cycles might also exist - which can be asymptotically free or safe. A fixed point can only be reached by trajectories within its UV critical surface, which is typically finite dimensional. Thus the requirement that a continuum limit exists eliminates an infinite number of free parameters from the model by imposing infinitely many relations that specify the location of the UV critical surface in theory space. 

However, what should the origin of similar relations between couplings be in the fundamentally discrete quantum- gravity setting?
The requirement that the effective, macroscopic dynamics takes a specific form is typically insufficient to uniquely determine the microscopic dynamics. This is due to infrared universality, i.e., the property that particular forms of the effective macroscopic dynamics act as attractors, and microscopic information is washed out during the coarse-graining procedure. In the case of quantum gravity, this property could become pivotal for the question of predictivity, as it might suggest that the requirement to recover General Relativity in the infrared limit could be too weak a condition to sufficiently constrain the microscopic dynamics. In the case of continuum approaches to quantum gravity, the requirement of a continuum limit is a strong condition, generically expected to render the model predictive. For physically discrete models, i.e., models with a physical ultraviolet cutoff, this requirement falls away, and an infinite possibility of choices for the microscopic dynamics might open up. 
In this setting, there might be the freedom to change the microscopic physics without resulting in (detectable) changes in the macroscopic physics. 
Note that this  is a potential challenge of any quantum- gravity model that aims to be fundamental, comes with a finite UV cutoff, and has an infinite number of couplings that are compatible with the symmetries and degrees of freedom of the model -- even if these are typically simply set to zero.

Crucially, the intuition about IR universality and the corresponding difficulty to restrict the UV dynamics based on the IR dynamics derives from a setting where the UV is local. Causal sets however are inherently non-local. This manifests itself in an infinite number of nearest neighbours for any element in a causal set that, e.g., is approximated by Minkowski spacetime. Moreover, in any given frame, most of the nearest neighbours are at arbitrarily large spatial distance, as they ``hug" the lightcone. 
Thus it is crucial to test whether the above intuition applies to this context. This requires to develop coarse- graining tools that allow to explore the theory space for causal sets. At the same time, the nonlocality  implies that the notion of coarse graining has to be adapted such that it can be applied to causal sets.   
\section{Coarse-graining for causal sets}\label{sec:floweq}
\subsection{Notion of coarse-graining in background-independent quantum- gravity models}
Traditionally, the notion of coarse-graining and the Renormalization Group is linked tightly to the existence of a background (Euclidean) spacetime. The reason is that one performs \emph{local} coarse graining, by averaging over high-energy degrees of freedom and thereby obtains an effective dynamics for the low-energy degrees of freedom. To define what one means by high-energy modes, one typically relies on a background metric that provides a generalized notion of momentum through the corresponding covariant Laplacian $\Delta$.  In the case of quantum gravity, if one insists on this local notion of coarse graining, one typically introduces a background metric, thus breaking background-independence, at least at intermediate steps of the procedure  --  physical quantities can still be independent of the choice of background, see, e.g., \cite{Becker:2014qya}. An alternative way to implement coarse-graining for quantum gravity is by ``averaging" over configurations with many degrees of freedom to obtain configurations with fewer degrees of freedom.
In particular, this notion has been employed to construct a Renormalization Group equation for matrix and tensor models, where the matrix size $N$ serves as the notion of scale in the path integral \cite{Brezin:1992yc,Eichhorn:2013isa,Eichhorn:2014xaa,Eichhorn:2017xhy}. This idea can also be applied to the group field theory setting \cite{Benedetti:2014qsa}.
Essentially, this setup allows to
 perform the path integral step wise, without resorting to the definition of a ``momentum-shell", as used in local coarse graining procedures. For causal sets, one instead orders fluctuations according to their position in the link matrix, and integrates over  its rows and columns successively instead of all at once in the path integral. In fact, heuristically the coarse graining proceeds just as one would expect, by thinning the number of discrete points and the corresponding causal connections.
 
In causal sets, one can introduce the size of the link matrix $N$ as an infrared cutoff, i.e., fluctuations in those elements $L_{ij}$ of the link matrix will be suppressed for which $i,j <N$. Thus, at first only the outermost elements of the link matrix are integrated out in the path integral. In a ``natural" labelling of the causal set elements, those are the elements that lie furthest to the future. One might thus picture the coarse- graining process as first ``averaging" over all causal sets that only differ in the links between the elements that are furthest to the future. Step by step one then averages over fluctuations in the causal set until finally all possible configurations have been averaged over.
The ultraviolet, microscopic limit is given by $N \rightarrow \infty$ (or $N \rightarrow N_{\rm max}$, if one works in a setting with a maximum number of elements, implementing, e.g., a universe with a finite Hubble volume), whereas the infrared corresponds to $N \rightarrow 0$, i.e., where quantum fluctuations in all matrix entries have been ``averaged" over.

Here, a first step towards developing a practical implementation of these ideas for causal sets is taken. As causal sets are intrinsically Lorentzian, and cannot be Wick-rotated, this is a procedure to coarse grain that is tied to Lorentzian quantum gravity.

\subsection{Flow equation for causal sets}
A  coarse-graining scheme for causal sets can be based on the size $N$ of the link matrix as cutoff scale. The construction is most straightforward for the analytical continuation of the source-term dependent path integral for causal sets
\be
Z[J] = \int \mathcal{D}L\, e^{-\alpha S[L] + \sum_{i,j}J_{ij}L_{ij}},
\ee 
where $\alpha = -i$ gives back the Lorentzian path integral. Note that the configurations in the path integral are \emph{not} rotated, as there is no meaningful way to discuss causality and construct a causal set in a Euclidean setting. The introduction of $\alpha$ allows to set up the coarse-graining procedure in close analogy to the procedure in matrix models, and also underlies Monte Carlo simulations of the path integral for causal sets \cite{Surya:2011du}. 
For a well-defined generating functional, one might start from the generating functional restricted to link matrices of maximum size $N_{\rm max}$. The integral is not performed all at once, but step-wise, by integrating out matrix elements down to an IR cutoff $N$, first, and then successively lowering $N$.

To that end one introduces a regulator term that allows to perform the path integral over the elements with $i,j>N$  of the link matrix first, producing a path integral where only the elements with $i,j<N$ still fluctuate. 
The scale-dependent generating functional is defined as
\be
Z_N[J] =\int \mathcal{D}L\, e^{- \alpha \left(S[L] - \sum_{i,j}J_{ij} L_{ij} + \Delta S_N[L]\right)}.
\ee
The regulator is chosen to be quadratic in $L_{ij}$, with a large amplitude for $i,j<N$: This suppresses the fluctuations in these components in the path integral. Thus, only the path integral over the fluctuations in the components $i,j>N$ is performed. Specifically, the regulator takes the form
\be
\Delta S_N[L]=\frac{1}{2}\sum_{ijkl}L_{ij} R_N (i,j,k,l) L_{kl}.
\ee
Essentially, this term mimicks the quadratic term in the action $ \sum_{i,j,k}L_{ij}L_{jk}$, but adds an index-dependence to the sum. 
One can thus understand the regulator as a simple, $N$ dependent modification of the two-point function.

The specific form of the cutoff function might be chosen, for instance, as follows:
\be
R_N(i,j,k,l) = \delta_{jk}\left(\frac{4N}{i+j+k+l}-1 \right)\theta(4N-i-j-k-l),\label{eq:reg}
\ee
For $N \rightarrow 0$, this vanishes. Accordingly the full path integral is recovered in that limit, i.e., $Z_N[J] \rightarrow Z[J]$, just as one demands in this limit where all quantum fluctuations should be included.
The choice in Eq.~\eqref{eq:reg} suppresses fluctuations in the components $i,j,k,l<N$. Thus, at finite $N$, only the ``UV modes" with indices larger than $N$ are integrated over, whereas the ``IR modes" remain for a later step.  The suppression is stronger for those modes with smallest index values.  This breaks the discrete version of diffeomorphisms, which are relabellings of the causal set elements. Next to a possible gauge-fixing term, the regulator is an additional source of gauge-symmetry breaking,  just as in the case of continuum gauge theories, see, e.g., \cite{Gies:2006wv}.

The derivation of the Wetterich equation \cite{Wetterich:1992yh} for continuum QFTs, that has been adapted to the pre-geometric matrix model case in \cite{Eichhorn:2013isa} can now be translated to  the setting of causal sets. The scale- dependent effective action is defined as
\be
\Gamma_N [\bar{L}]= \sum_{ij}J_{ij}\bar{L}_{ij} -{\rm ln} Z_N- \Delta S_N[\bar{L}],
\ee
and depends on the ``classical" link matrix $\bar{L}$, which corresponds to the expectation value, i.e., 
\bea
\bar{L}_{ij} = \langle L_{ij} \rangle_N = \frac{1}{Z_N[0]}\int \mathcal{D}L\, L_{ij}\, e^{- \left( S[L] + \Delta S_N[L]\right)}=\frac{d {\rm ln}Z_N}{d J_{ij}}\Big|_{J=0}.\label{eq:defclassicalL}
\eea
Accordingly,
\bea
J_{ij}& =& \frac{d\Gamma_N[\bar{L}]}{d \bar{L}_{ij}}+ \sum_{k,l} R_N(i,j,k,l)\bar{L}_{kl}.
\eea
From this, one concludes that 
\bea
\frac{dJ_{ij}}{d\bar{L}_{kl}}= \frac{d^2\Gamma_N[\bar{L}]}{d\bar{L}_{ij}\, d\bar{L}_{kl}}+ R_N(i,j,k,l).\label{eq:dJdL}
\eea
Further, one can derive from eq.~\ref{eq:defclassicalL} that
\bea
\frac{d\bar{L}_{ij}}{dJ_{kl}}&=& \langle L_{ij} \, L_{kl}\rangle_N - \bar{L}_{ij} \bar{L}_{kl}.\label{eq:dLdJ}
\eea
Now, the unit operator can be re-expressed as
\be
\delta_{im}\delta_{jn} = \frac{d L_{ij}}{dL_{mn}}= \frac{dL_{ij}}{dJ_{kl}}\frac{dJ_{kl}}{dL_{mn}},
\ee
and the relations \eqref{eq:dJdL} and \eqref{eq:dLdJ} can be used.
This yields
 \be
 \sum_{k,l}\left(\langle L_{ij} \, L_{kl}\rangle_n - \bar{L}_{ij} \bar{L}_{kl}\right)\left(\frac{d^2\Gamma}{dL_{kl}dL_{mn}}+R_{N}(k,l,m,n) \right)= \delta_{im}\delta_{jn}.\label{eq:GGammainv}
 \ee
 Accordingly, the propagator of the theory is given by $\left(\frac{d^2\Gamma}{dL_{kl}dL_{mn}}+R_{N}(k,l,m,n)\right)^{-1}$.
One can now use Eq.~\eqref{eq:GGammainv} to derive an equation for the scale-dependence of $\Gamma_N$:
\bea
\partial_N \Gamma_N[\bar{C}]&=& - \partial_N {\rm ln}Z_N - \frac{1}{2}\bar{L}_{ij}\partial_N R_N(i,j,k,l)\bar{L}_{jk}
\nonumber\\
&=&-\frac{1}{Z_N} \int \mathcal{D}L\, e^{-\left(S[L]- \sum_{ij}J_{ij} L_{ij} +\Delta S_N[L] \right)}\left(-\frac{1}{2}\right)L_{ij}\partial_N R_N(i,j,k,l)L_{jk}\nonumber\\
&{}& -\frac{1}{2}\bar{L}_{ij}\partial_N R_N(i,j,k,l)\bar{L}_{jk}\nonumber\\
&=&\sum_{i,j,k,l}\left(\left(\langle L_{ij} \, L_{jk}\rangle_N - \bar{L}_{ij} \bar{L}_{jk} \right)\frac{1}{2}\partial_N R_N(i,j,k,l)\right)\nonumber\\
&=& \sum_{i,j,k,l}\partial_N R_N(i,j,k,l)\left( \frac{d^2\Gamma_N[\bar{L}]}{d\bar{L}_{ij}\, d\bar{L}_{kl}}+ R_N(i,j,k,l)\right)^{-1}.\label{eq:floweq}
\eea
Note that this equation is exact, as no approximations enter its derivation.

For practical applications, approximations must be introduced: $\Gamma_N$ contains all infinitely many operators compatible with the symmetries, and thus this flow equation corresponds to an infinite tower of coupled differential equations for the $N$-dependent couplings of the model. Solving it requires a truncation, i.e., picking a subspace of the space of couplings.

In gauge theories one generally cannot directly invert $\Gamma^{(2)}$, as gauge invariance results in zero-modes in the inverse propagator, requiring a gauge-fixing in order to derive the (scale-dependent) propagator.  If the label invariance of the theory space has a similar consequence here, a gauge fixing term can be introduced which picks one representative of each gauge orbit, e.g., by enforcing a form of the link matrix that satisfies the condition of natural labelling, where its lower triangular part contains only zeros.

The flow equation \eqref{eq:floweq} could allow to make progress towards answering the questions raised in the introduction: 

Analyzing the flow equation in truncations of the space of couplings, e.g., starting with the couplings contained in the Benincasa-Dowker action, could help to elucidate whether this action dynamically suppresses non-manifoldlike configurations such that a smooth spacetime emerges in the IR, or whether further terms, e.g., such that correspond to discrete counterparts of higher-order curvature operators, need to be added to achieve this. The flow equation derived here is a reformulation of the path integral. As such, finding its solution is equivalent to performing the path integral, but is technically different, and might therefore offer advantages in its implementation. It might also be worthwhile to test whether a similar flow equation in terms of the causal matrix might work better. While the full path integral based on the two should be completely equivalent, truncations might converge faster in one formulation than in the other.

The causal set theory space does not come endowed with a natural notion of scaling dimension. However, the use of $N$ as a scale implies that couplings should acquire a ``canonical" scaling with $N$.
This is analogous to the case of matrix and tensor models, where it has been shown how the flow equation automatically determines a scaling dimension of couplings with $N$ if the requirement of a well-defined large-$N$ limit is imposed \cite{Eichhorn:2017xhy}. A similar procedure should be applicable to the case of causal sets. This is the first step towards analyzing whether the intuition from the local continuum case carries over, as it will show whether the causal set theory space has an analogous ordering principle according to ``canonical" dimensionality.

Addressing the search for asymptotic safety requires to search for fixed points in the limit $N \rightarrow \infty$. 
In this interpretation, the discreteness scale is understood as a technical tool to get a handle on the path integral, instead of a physical scale. Thus, the causal- set framework could be reinterpreted as a possible way to discretize the path integral for quantum gravity. When the discreteness scale is introduced as a technical tool instead of a physical discreteness scale, physics is encoded in a universal continuum limit that is independent of the microscopic details. Rephrased in RG language this corresponds to a fixed point, at which universality is tied to scale-invariance.
If found, a comparison of its critical exponents to those of the continuum case could shed light on the universality class and a possible connection to continuum studies.

\acknowledgements{I thank L.~Glaser and S.~Surya for discussions. I acknowledge funding by the DFG within the Emmy-Noether-program under grant no.~Ei-1037-1 and support by the Perimeter Institute for Theoretical Physics through the Emmy-Noether-visiting fellow program. }

\end{document}